\documentclass[conference]{IEEEtran}
\renewcommand{\baselinestretch}{1.0}
\usepackage{booktabs} 
\usepackage{epsfig}
\usepackage[TABBOTCAP]{subfigure}
\usepackage{tabularx}
\usepackage{graphicx}
\usepackage{color}
\usepackage{xspace}
\usepackage{thumbpdf}
\usepackage{listings}
\usepackage{verbatim}
\usepackage{booktabs}
\usepackage{colortbl}
\usepackage{url}
\usepackage{nth}
\usepackage{subfigure}
\usepackage{caption} 
\usepackage{multirow}
\usepackage{multicol}

\begin{document}
%
\title{Adversarial Attacks on Cognitive Self-Organizing Networks: The Challenge and the Way Forward}

\author{\IEEEauthorblockN{Muhammad Usama}
\IEEEauthorblockA{Information Technology University\\
Punjab, Pakistan\\
muhammad.usama@itu.edu.pk}
\and
\IEEEauthorblockN{Junaid Qadir}
\IEEEauthorblockA{Information Technology University\\
Punjab, Pakistan\\
junaid.qadir@itu.edu.pk}
\and
\IEEEauthorblockN{Ala Al-Fuqaha}
\IEEEauthorblockA{Western Michigan University, USA\\
ala.al-fuqaha@wmich.edu}}

\maketitle

\begin{abstract}
Future communications and data networks are expected to be largely cognitive self-organizing networks (CSON). Such networks will have the essential property of cognitive self-organization, which can be achieved using machine learning techniques (e.g., deep learning). Despite the potential of these techniques, these techniques in their current form are vulnerable to adversarial attacks that can cause cascaded damages with detrimental consequences for the whole network. In this paper, we explore the effect of adversarial attacks on CSON. Our experiments highlight the level of threat that CSON have to deal with in order to meet the challenges of next-generation networks and point out promising directions for future work.
\end{abstract}

\IEEEpeerreviewmaketitle
\section{Introduction}
\sloppy
The idea that networks should learn to drive themselves is gaining traction \cite{feamster2017and}, taking inspiration from self-driving cars where driving and related functionality do not require human intervention. The networking community wants to build a similar cognitive control in networks where networks are able to configure, manage, and protect themselves by interacting with the dynamic networking environment.We refer to such networks as cognitive self-organizing networks \textit{CSON}. The expected complexity and heterogeneity of CSON makes machine learning (ML) a reasonable choice for realizing this ambitious goal. Recently artificial intelligence (AI) based CSON have attained a lot of attention in industry and academia.   

In 2003,  Clark et al. \cite{clark2003knowledge} proposed that ML and cognitive techniques should be used for operating the network, this knowledge plane incorporation will bring many advantages in networks such as automation of network management, efficient and real-time anomaly and intrusion detection and many related tasks. Due to limited computational resources and lack of ML abilities, the idea of knowledge plane was not implemented in networks. In recent years, the field of ML, especially ``neural networks'', have evolved rapidly and we have witnessed its success in vision, speech, and language processing. This huge success motivated networking research community to utilize deep ML tools for building CSON.

Deep ML or deep learning (DL) is a branch of ML where hierarchical architectures of neural networks are used for unsupervised feature learning and these learned features are then used for classification and other related tasks. DL classifiers are function approximators that require a lot of data for generalization. Although they have outperformed all other statistical approaches on large datasets, due to generalization error, they are very vulnerable to adversarial examples. Adversarial examples are carefully crafted perturbations in the input which when ML/DL algorithms are subjected to get classified in a different class with high probability.

In this paper, we take security to encompass the securing of all of the functional areas of CSON (i.e., ISO defined functional areas often abbreviated as FCAPS: fault, configuration, accounting, performance, and security) and experiment with multiple adversarial attacks on ML/DL based malware classification systems. Our experimental results demonstrate that the current state of the art ML/DL based networking solutions do not have substantial deterrence against adversarial attacks. Specifically, our experiments utilize the highly cited malware image dataset provided by Nataraj et al. \cite{nataraj2011malware} to perform adversarial attacks on malware classifier to demonstrate that using current ML/DL techniques in conjunction with CSONs can be a potential security risk.

\textbf{Contributions}:
In this paper, we have made the following contributions:
\begin{itemize}
  
    \item To the best of our knowledge, we have made the first attempt to show that CSON utilizing ML/DL techniques are very vulnerable to attacks based on adversarial perturbations.
  
    \item We have argued that existing defenses to overcome adversarial perturbations are not appropriate and efficient for CSON applications. We have also highlighted that protection schemes against adversarial examples create an arms race between adversaries. 
    
\end{itemize}

The rest of the paper is organized as follow. In the next section, we review related research studies that focus on CSON and adversarial attacks on networking applications. Section III describes our research methodology; particularly, with reference to the dataset, the ML/DL model, used dataset, and threat model assumptions, and the adversarial attacks. In Section IV, we provide the details of our experimental evaluations and the potential defense against these attacks. In Section V, we discuss the posed questions as well as some future directions and challenges. Finally, Section VI concludes our study.

\section{Related Work}
Many applications of ML/DL in networking have been proposed in the last few years highlighting the applications, opportunities, and challenges of using ML/DL in networking domain \cite{alsheikh2014machine, bartulovic2017biases, bkassiny2013survey, fadlullah2017state, hodo2017shallow,
klaine2017survey, patcha2007overview, usama2017unsupervised, zhang2018deep, wang2018machine}.  

Although many ML-based solutions for networking applications have been proposed, the networking community has not yet standardized any ML-based solutions for CSONs. This arises partly from the complexity of the CSON environment that is characterized by dynamically changing network environment, data sparsity, expected tussles between control loops, high dimensionality, label data scarcity, heterogeneity, offline data processing, and many other architectural issues.

CSON are expected to resolve the challenges of optimization, configuration, healing, and coordination in the communication and data networks by incorporating AI/ML based cognitive techniques. Latif et al. \cite{latif2017artificial} highlights AI as a potential enabler for CSON. Similar ideas based on deep reinforcement learning for learning from environment and experience termed as experience-driven networking are presented in \cite{xu2018experience}. Feamster et al. \cite{feamster2017and} termed this idea of learning from network environment for measuring, analyzing, and configuring network without any human intervention as self-driving networks. Jiang et al. \cite{jiang2017unleashing} highlighted the benefits and challenges in developing an intelligent data-driven network with the ability of learning from dynamic nature of the networking environment by using exploration and exploitation processes. Koley et al. \cite{koley2016zero} proposed and provided a framework for zero-touch networking and highlighted the need for CSON using Google's infrastructure network as an example. Mestres et al. \cite{mestres2017knowledge} revisited the possibilities of embedding artificial intelligence in networking and proposed an ML/DL based knowledge plane for networking applications and this new networking paradigm was termed as knowledge defined networking.  


While ML/DL applications will be a core part of CSON, recent studies demonstrated that ML/DL models are very susceptible to adversarial examples \cite{liu2018survey} \cite{akhtar2018threat}. Although most existing studies in this domain have targeted image classification applications in which high-dimensional images are perturbed in a way that fools the algorithm without being the change being conspicuous to naked human eye, these attacks also pose a significant challenge to CSON since the underlying algorithms are largely similar. 

Such adversarial attacks are performed to compromise the integrity in terms of misclassification, accuracy reduction, targeted misclassification, or decision boundary evasion of the ML/DL techniques. We can divide these adversarial attacks into two broader categories based on the adversary's/attacker's knowledge.
\begin{itemize}
    \item \textbf{White-box Attack}: This attack assumes that the adversary has complete knowledge about the ML/DL architecture, training data, and hyper-parameters. For adversarial attacks on CSON, we assume a white-box attack setting.
    \item \textbf{Black-box Attack}: This attack assumes that the adversary/attacker has no information about the ML/DL technique and hyper-parameters. The adversary acts as a standard user who can query the ML/DL based system and gets a response. These query-response pairs are later used for crafting the adversarial examples. 
\end{itemize}

Most of the adversarial attacks are white-box attacks, but white-box adversarial examples can be converted into black-box attacks by exploiting the ML/DL transferability property \cite{papernot2016transferability}. 

Since these adversarial attacks on ML algorithms have not yet been applied much in the case of networks, we will initially review their applications in other domains. Szegedy et al. \cite{szegedy2013intriguing} proposed the first successful adversarial attack that has fooled the state of the art image classifiers with very high probability. Goodfellow et al. \cite{goodfellow2014explaining} proposed an adversarial sample generation method called \textit{fast gradient sign method}, where adversarial perturbation was generated by taking the sign of the gradient of the cost function with respect to the input. Kurakin et al. \cite{kurakin2016adversarial} explored the vulnerability of ML/DL techniques in the physical world and demonstrated that a small invisible tweak to the input of an ML/DL techniques can result in incorrect results. Carlini et al. \cite{carlini2017towards} proposed three attacks by exploiting the three different distance matrices $(L_0, L_2,$ and $L_\infty)$ and showed that the defensive distillation method \cite{papernot2016distillation} used to prevent against adversarial attacks does not increase the robustness of the ML/DL techniques. Papernot et al. \cite{papernot2016limitations} proposed a \textit{saliency map based attack}, where saliency map is used to find the most discriminative features of the input that are then fractionally perturbed to form an adversarial attack on the ML/DL based classifiers. In 2017, Papernot et al. \cite{papernot2017practical} proposed a black-box attack, where adversarial attack transferability \cite{papernot2016transferability} is exploited to form a successful evasion attack. Further details about adversarial attacks on different vision, language, and text processing systems can be found in \cite{yuan2017adversarial} and \cite{akhtar2018threat}.

Adversarial attacks have not yet been explored for CSON, we will cover some general networking applications. In 2013, Corona et al. \cite{corona2013adversarial} highlighted the possibilities and open research challenges of adversarial attacks on intrusion detection systems. Hu et al. \cite{hu2017generating} proposed a generative adversarial network (GAN) based black-box attack on malware examples but training a GAN on malware examples is difficult and computationally exhaustive. Grosse et al. \cite{grosse2016adversarial} proposed an adversarial perturbation attack against deep neural networks for malware classification, where a restricted amount of feature perturbations are used to fool a deep neural network with $0.85$ probability which was previously classifying malware with $97\%$ accuracy. In the next section, we provide the details of the proposed approach to perform multiple adversarial attacks on CSON.

\section{Methodology}
In this section, we describe the approach followed in designing adversarial examples to evade the ML/DL based malware classification system which we use as a proxy for the functional areas of CSON. To the best of our knowledge, no standardized deep learning based solution for malware classification in the CSON has been proposed yet. In this work, we propose a deep neural network based solution for malware classification. Before delving deep into the details of the proposed model, we describe the threat model and few related assumptions. 
\subsection{Threat Model}
In the following, we outline the salient assumptions regarding the adversarial threat:
\begin{itemize}
\item The adversary may have the knowledge about the trained model which includes model architecture and hyper-parameters, but the adversary cannot make any changes to the architecture or model parameters. This is a common assumption in the adversarial machine learning domain \cite{akhtar2018threat}.
\item The adversary can only perform attacks during the testing phase, attacks on the training data (i.e., poisoning attacks) are not within the scope of this study. 
\item For malware classification, we assume that similar families of malware, when represented as grayscale images exhibit similar visual and texture representations. This hypothesis was proposed and defended in \cite{nataraj2011malware}. In this work, we utilize convolutional neural networks (CNN) for malware classification because CNN is by far the best feature extractors. 
\item The goal of an attack is to compromise the integrity of the ML/DL based classification techniques through a reduction in the classification accuracy with small perturbations. 
\end{itemize}

\subsection{Malware Image Representation}
In this paper, we have used grayscale malware image dataset provided in \cite{nataraj2011malware}, where a malware executable is converted to a grayscale image. This approach of conversion includes both static and dynamic code analysis. The executable code is converted to binary and then represented as 8-bit unsigned vectors, these 8-bit unsigned vectors are then reshaped to a 2D array which can be visualized as a grayscale image. Figure 1 is depicting the procedure of converting malware executable to a grayscale image.   
\begin{figure} 
\centering     
\centerline{\includegraphics[width=0.5\textwidth]{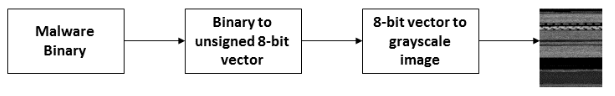}}
\caption{Depiction of malware executable as an image}
\end{figure}
\subsection{Malware Classification Model}
We propose a CNN based malware classification architecture. Table \ref{ta1} depicts the proposed architecture. CNN is a powerful DL technique that learns spatial feature representations using convolutional filters. CNN has the capability to tolerate the distortion and spatial shifts in the input data and extract features from raw input data. CNN provides the state-of-the-art solution for network traffic feature extraction and classification \cite{lotfollahi2017deep}, motivated by these successes, we explore the use of CNN for grayscale malware image classification. 

In the proposed architecture, we re-scale the input grayscale images of various sizes to $28$ pixel wide and $28$ pixel high, where pixel values are between $0$ to $255$. These input values are subjected to a two-dimensional convolutional layer with $64$ filters of receptive field $14$ pixel wide and $14$ pixel high. After that, we use a rectified linear unit (i.e., ReLU) as an activation function. The resultant activation values are then passed on to a second convolution layer with $128$ filters and $5\times5$ receptive field. Again, we use a ReLU as an activation function. Similarly, the third convolution layer follows the same procedure mentioned earlier but with $128$ filters of $1\times1$ receptive field. After the third convolution layer, the resultant activation values are flattened and passed on to a fully connected layer with softmax as an activation function producing resulting probabilities. We use a variant of the stochastic gradient descent (SGD) as an optimization function and categorical cross-entropy as a loss function to train the CNN.  

\begin{table}[!ht]
\centering
\scriptsize
\caption{Proposed CNN architecture for malware classification}
\label{ta1}
\begin{tabular}{|c|}
\hline
\begin{tabular}[c]{@{}c@{}}\textbf{Input}: Malware gray scale image,\\           Size: (28*28)\end{tabular}                                       \\ \hline
\begin{tabular}[c]{@{}c@{}}\textbf{2D Convolution Layer}\\ (Filter Size: 14*14,\\  No. of filters: 64, \\ Activation Function: ReLU)\end{tabular} \\ \hline
\begin{tabular}[c]{@{}c@{}}\textbf{2D Convolution Layer}\\ (Filter Size: 5*5,\\ No. of filters: 128,\\ Activation\\ Function: ReLU)\end{tabular}  \\ \hline
\begin{tabular}[c]{@{}c@{}}\textbf{2D Convolution Layer}\\ (Filter Size: 1*1,\\ No. of filters: 128,\\ Activation\\ Function: ReLU)\end{tabular}  \\ \hline
\begin{tabular}[c]{@{}c@{}}\textbf{Dense Layer} \\ (Number of neurons: 25,\\ Activation function: Softmax)\end{tabular}                           \\ \hline
\begin{tabular}[c]{@{}c@{}}\textbf{Output}: \\ Malware\\ Classification Probabilities\end{tabular}                                                \\ \hline
\end{tabular}
\end{table}

\subsection{Adversarial Attacks}\label{adv}
We performed \textit{fast gradient sign method}, \textit{basic iterative method}, and \textit{Jacobian-based saliency map} attacks on malware classifiers to demonstrate that ML/DL based malware classification methods in CSON are vulnerable to adversarial examples.

\subsubsection{Fast Gradient Sign Method}
Goodfellow et al. \cite{goodfellow2014explaining} proposed a fast method of generating adversarial examples, this method is called the fast gradient sign method (FGSM). This method exploits the vulnerability of deep neural networks to adversarial perturbations. FGSM performs one step gradient update along the sign of the gradient to solve the optimization problem. Formally, the perturbation is calculated as:
\begin{equation}\label{eq1}
\eta =\epsilon \textit{sign}(\nabla_x j_\theta(x,l))
\end{equation}
In equation \ref{eq1}, $\epsilon$ represents the update step width or magnitude of the perturbation, $\eta$ is the difference between original and perturbed input, $\nabla_x$ represents the gradient with respect to each example, lastly $j_\theta(x,\textit{l})$ is the loss function used for training the neural network for original example $x$ and its corresponding label $\textit{l}$.
The generated adversarial example $x^{'}$ is calculated as: 
\begin{equation}\label{eq2}
x^{'} = x + \eta 
\end{equation}
FGSM is a very powerful attack because it is resilient to the regularization techniques such as dropout and norm-based regularization methods. 
\subsubsection{Basic Iterative Method}
Kurakin et al. \cite{kurakin2016adversarial} proposed an element-wise basic iterative method (BIM) for adversarial falsification. It is an iterative procedure for generating adversarial example for physical world applications. They improved the success rate of the FGSM attack by including an iterative clipping method for each pixel to avoid large changes in the pixel values.The generated adversarial example is calculated via multiple iterations. The adversarial example generation procedure is given as:
\begin{equation}\label{eq3}
x_{0} = x,
\end{equation}
\begin{equation}\label{eq4}
x_{n+1} = \textit{Clip}{_x,_\xi}({{x{_n} + \epsilon \textit{sign}(\nabla_x j_\theta(x,l))}})
\end{equation}
Where $x_{n+1}$ is an adversarial example after $n+1$ iterations. The rest of the parameters are similar to the one utilized in the FGSM attack. 
\subsubsection{Jacobian-based Saliency Map Attack}
Papernot et al. \cite{papernot2016limitations} proposed a new efficient method for generating adversarial examples called the Jacobian-based saliency map attack (JSMA). This attack is an iterative method for generating a saliency map to find out the most discriminative features, a Small perturbation is added to these discriminative features to fool the classifier. This attack is based on calculating the Jacobian of the forward propagating examples with respect to the input sample. The procedure of generating the saliency map of each sample is given as: 
\begin{equation}\label{eq5}
J(x) = \frac{\partial f(x)}{\partial x} = [\frac{\partial f{_j}(x)}{\partial (x{_i})}] 
\end{equation}
This attack achieved $97\%$ accuracy by altering only $4.2\%$ of the input features. Although this attack provides very effective adversarial examples but it is computationally very expensive \cite{papernot2016limitations}. 

\section{Experimental Evaluation} \label{ee}
We evaluated the CNN based malware classifier against adversarial examples. Through our experiments, we want to answer the following questions:
\begin{itemize}
\item \textbf{Question 1:} \textit{Since ML/DL techniques are necessary to fuel the CSON, do these techniques provide the necessary robustness required to deal with adversarial perturbations?}
\item \textbf{Question 2:} \textit{How to build deterrence against adversarial attacks in CSON?}
\item \textbf{Question 3:} \textit{Do the deterrence techniques against adversarial examples create an arms race between adversaries?}
\end{itemize}
Before answering these questions, we provide the details of the dataset used for our experiments.
\subsection{Dataset}\label{data}
Nataraj et al. \cite{nataraj2011malware} provided a malware grayscale images dataset based on their novel image processing technique where malware execute-able are viewed as a grayscale image for visualizing malware families for classification purposes. We evaluated the performance of our proposed CNN architecture and adversarial attacks on malware classifiers using this dataset. The dataset consists of $9,458$ malware images divided into $25$ different malware families like Allaple.L, Allaple.A, Lolyda. AA etc. These malware families belong to major malware types such as worm, PWS, trojan, Dialer, Tdownloader, rouge, and backdoor, more details about malware types and related families in the dataset is available in \cite{nataraj2011malware}. Here we want to highlight that to keep the excutability of the malware we have limited the scope of the perturbation to the uninitialized data and zero padding portion of the malware image. We utilized $70\%$ of the data for training and $30\%$ for testing. Figure 2 depicts a sample malware image and its associated attributes.

\begin{figure}
\centering     
\centerline{\includegraphics[width=0.25\textwidth]{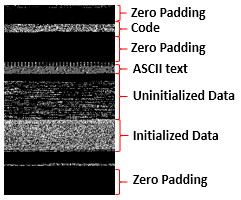}}
\caption{Malware image and related features in the image.}
\end{figure}
\begin{table*}[]
\centering
\tiny
\label{table2}
\caption{FGSM attack and defense results with different values of epochs and $\epsilon$}
\begin{tabular}{|c|l|c|c|c|}
\hline
\multicolumn{5}{|c|}{Fast Gradient Sign Method Attack} \\ \hline
\multicolumn{1}{|l|}{Epochs} & Epsilon & \begin{tabular}[c]{@{}c@{}}Test accuracy on\\  Legitimate Samples in (\%)\end{tabular} & \begin{tabular}[c]{@{}c@{}}Test accuracy of \\ Adversarial Examples in (\%)\end{tabular} & \begin{tabular}[c]{@{}c@{}}Test accuracy after\\  Adversarial training in (\%)\end{tabular} \\ \hline
\multirow{3}{*}{10} & 0.1 & 98.39 & 1.87 & 78.16 \\ \cline{2-5} 
 & 0.2 & 98.29 & 0.37 & 88.17 \\ \cline{2-5} 
 & 0.3 & 97.97 & 0.7 & 91.7 \\ \hline
\multirow{3}{*}{50} & 0.1 & 98.02 & 1.61 & 94.91 \\ \cline{2-5} 
 & 0.2 & 98.72 & 0.48 & 98.07 \\ \cline{2-5} 
 & 0.3 & 97.97 & 0.27 & 97.75 \\ \hline
\multirow{3}{*}{100} & 0.1 & 98.34 & 1.55 & 96.73 \\ \cline{2-5} 
 & 0.2 & 97.34 & 0.32 & 96.6 \\ \cline{2-5} 
 & 0.3 & 97.64 & 0.32 & 97.43 \\ \hline
\end{tabular}
\end{table*}
\begin{table*}[]
\centering
\tiny
\caption{BIM attack and defense results with different values of epochs and $\epsilon$}
\label{tab3}
\begin{tabular}{|c|l|c|c|c|}
\hline
\multicolumn{5}{|c|}{Basic Iterative Method Attack} \\ \hline
\multicolumn{1}{|l|}{Epochs} & Epsilon & \begin{tabular}[c]{@{}c@{}}Test accuracy on\\  Legitimate Samples in (\%)\end{tabular} & \begin{tabular}[c]{@{}c@{}}Test accuracy of \\ Adversarial Examples in (\%)\end{tabular} & \begin{tabular}[c]{@{}c@{}}Test accuracy after\\  Adversarial training in (\%)\end{tabular} \\ \hline
\multirow{3}{*}{10} & 0.1 & 97.22 & 0.90 & 61 \\ \cline{2-5} 
 & 0.2 & 96.95 & 0.75 & 36 \\ \cline{2-5} 
 & 0.3 & 97.32 & 0.91 & 35 \\ \hline
\multirow{3}{*}{50} & 0.1 & 97.91 & 0.48 & 72 \\ \cline{2-5} 
 & 0.2 & 98.07 & 0.27 & 38 \\ \cline{2-5} 
 & 0.3 & 97.38 & 0.70 & 30 \\ \hline
\multirow{3}{*}{100} & 0.1 & 97.91 & 0.70 & 77 \\ \cline{2-5} 
 & 0.2 & 98.81 & 0.64 & 47 \\ \cline{2-5} 
 & 0.3 & 97.59 & 1.02 & 31 \\ \hline
\end{tabular}
\end{table*}
\begin{table*}[]
\centering
\tiny
\caption{JSMA attack with average number of features perturbed for different values of epochs and $\gamma$}
\label{tab4}
\begin{tabular}{|c|l|c|c|c|}
\hline
\multicolumn{5}{|c|}{Jacobian-based Saliency Map Attack} \\ \hline
\multicolumn{1}{|l|}{Epochs} & Gamma & \begin{tabular}[c]{@{}c@{}}Test accuracy on\\  Legitimate Samples in (\%)\end{tabular} & \begin{tabular}[c]{@{}c@{}}Test accuracy of \\ Adversarial Examples in (\%)\end{tabular} & \begin{tabular}[c]{@{}c@{}}Average number of\\  Features Perturbed (\%)\end{tabular} \\ \hline
\multirow{3}{*}{10} & 0.1 & 95.93 & 9.13 & 90.5 \\ \cline{2-5} 
 & 0.2 & 97 & 4.8 & 95.8 \\ \cline{2-5} 
 & 0.3 & 96.62 & 4.8 & 95.2 \\ \hline
\multirow{3}{*}{50} & 0.1 & 97.53 & 4.83 & 94.28 \\ \cline{2-5} 
 & 0.2 & 97.48 & 5.0 & 91.04 \\ \cline{2-5} 
 & 0.3 & 96.62 & 4.8 & 95.2 \\ \hline
\multirow{3}{*}{100} & 0.1 & 98.28 & 19.67 & 75.88 \\ \cline{2-5} 
 & 0.2 & 98.28 & 7.83 & 88.09 \\ \cline{2-5} 
 & 0.3 & 97.21 & 5.0 & 90.83 \\ \hline
\end{tabular}
\end{table*}
\subsection{Results}\label{sec4}
We evaluated the performance of adversarial attacks on CSON using malware classifiers as a proxy. The dataset details are provided in section \ref{data}. Both FGSM and BIM attacks are element-wise attacks, with individual perturbation scope, non-targeted specificity and same perturbation magnitude parameter $\epsilon$. We performed both attacks using multiple values of $\epsilon$ with $10$, $50$ and $100$ epochs. Our experimental results are shown in Tables II and III. JSMA is a targeted, iterative, Euclidean distance based attack. It has two major controlling parameters; namely, \textit{maximum distortion parameter} $\gamma$ and \textit{rate of perturbation in the features} $\theta$. For this experiment, we fixed $\theta$ to be $+1$ and varied the value of $\gamma$ between $0.1$, $0.2$ and $0.3$ for $10$, $50$ and $100$ epochs. The achieved adversarial test accuracy values along with the average number of features perturbed for a successful adversarial example are reported in Table \ref{tab4}. For all aforementioned experiments, a batch size of $128$ and a learning rate of $0.001$ were used.
\subsubsection{Performance impact}
The CNN based malware classifier has a classification accuracy of $98.39\%$ when trained on legitimate examples. This accuracy is better than the best accuracy reported on the dataset in consideration. Adversarial test examples created by employing FGSM have reduced the classification accuracy from approximately $99\%$ to $1.87\%$ which is nearly $97\%$ loss in the accuracy of classification and prevention against adversarial examples.  It also means that the probability of an adversary evading the malware classifier has increased from $1\%$ to $97\%$ which is very alarming. Similarly, the BIM attack reduces the test accuracy of adversarial samples to $0.9\%$ which is even worse than the FGSM attack. In case of JSMA, the classification accuracy decreased from $98.28\%$ to $7.87\%$ but it requires an $88.09\%$ of average feature perturbations to create successful adversarial examples, which is computationally very expensive. The full experimental results are summarized in Tables 2, \ref{tab3} and \ref{tab4}.

Malware classifiers are an integral part of the security architecture of CSON and we demonstrated that a very small perturbation in the test example has the potential to evade the integrity of the classifier. This performance degradation depicts the potential risks of applying ML/DL methods in the context of CSON without considering the robustness of ML/DL classifiers and building proper deterrence against adversarial examples. Without such deterrence, ML/DL models might cause more harm than good in CSON.   


\subsubsection{Computational complexity}
Adversarial attacks are not just random noise/values added to the test samples. Instead, they are carefully calculated perturbations. These perturbations are based on exploiting the inherent generalization error and gradient variations in of ML/DL techniques. As the shown in Table \ref{tab4}, detecting and exploiting these errors to make effective adversarial examples is a computationally very complex and expensive process. Since JSMA works on saliency maps and forward derivatives to find the most discriminant features, it becomes computationally very expensive. Table \ref{tab4} depicts the average number of features perturbed to construct an adversarial example for each class, these values are surprisingly very high because for each example the underlying data contains $784$ features and each feature has a value greater than zero which is not the case in other standard datasets like MNIST \cite{lecun2010mnist}. This unusual property of the malware image dataset increases the search space to find the most discriminating features; thus, resulting in rapid increase in that computational complexity and poor performance of the JSMA attack.

\subsection{Adversarial Defense}
We need to identify that adversarial settings have been assumed in networks before through tools such as game theory, but unique challenges emerge and the stakes get higher when we give more control of the network to ML and algorithms in CSON \cite{manshaei2013game}. Barreno et al. \cite{barreno2006can} provided a taxonomy of defences against adversarial attacks, they have highlighted that regularization, randomization and information hiding can ensure defence against adversarial perturbation but these countermeasures are not very effective against attacks described in section \ref{adv}.

There are two major types of defenses against adversarial examples; namely, \textit{proactive} and \textit{reactive}. Proactive defenses include \textit{adversarial training} and \textit{network distillation}. Whereas reactive defenses include \textit{input reconstruction} and \textit{adversarial detection}. In this paper, we only consider proactive countermeasures against adversarial examples. More detail about reactive countermeasures against adversarial examples are explored in \cite{barth2010learning}.     
\subsubsection{Adversarial Training}
One countermeasure against adversarial examples is to include adversarial examples in the training data for ML/DL techniques. Goodfellow et al. \cite{goodfellow2014explaining} proposed this idea and showed that ML/DL classifiers can be made more robust against adversarial examples by training them with adversarial examples. The purpose of including adversarial examples in the training is to regularize the ML/DL technique. This regularization helps to avoid over-fitting which in turn increases the robustness of the ML/DL technique against adversarial examples.

In this paper, we also explored adversarial training for making CNN models robust against FGSM and BIM attacks. Test accuracies before and after the adversarial training are reported in Tables II and \ref{tab3}. The results clearly show that performing adversarial training can increase the deterrence against adversarial attacks but it only provides defense against the adversarial examples on which it is trained, while other adversarial perturbations continue to pose a threat of evading the integrity of the classifier.  
\subsubsection{Network Distillation}
Network distillation is another approach of forming a defense against adversarial examples. Hinton et al. \cite{hinton2015distilling} proposed the idea of distillation to improve the generalization of the deep neural networks. Papernot et al. \cite{papernot2016distillation} used the distillation process to form a defense against adversarial examples. Network distillation is a process of training a classifier such that the generation of adversarial examples becomes very difficult. This defense is based on hiding the gradients between pre-softmax layers and the softmax output, which reduces the chances of developing a gradient-based attack against deep neural networks. Since in this paper we consider white-box attacks where an adversary knows the model parameters (i.e., architecture, hyper-parameters, gradients, etc.), this defensive scheme is not applicable to our study. More information on defence schemes against adversarial examples can be found in \cite{yuan2017adversarial}. 
\section{Discussions, challenges and Future Extensions} 
Our experimental results clearly demonstrate that applying ML/DL techniques in CSON without taking into account adversarial perturbation threats can potentially lead to major security risks. To date, there does not exist any appropriate solution that provides deterrence against all kinds of adversarial perturbations. Our experiments answer the questions posed earlier in Section \ref{ee}. Furthermore, they provide the following insights:
\begin{itemize}
    \item \textbf{Robustness of ML/DL for CSON}: In section \ref{sec4}, we have shown that CSON are very vulnerable to adversarial attacks. Sparsity, high dimensionality, unstructured nature, unique data packing scheme, large salient feature decision space of network data and less fault tolerance makes adversarial attacks more lethal for CSON as compared to other vision and language data. Given the adversarial threat, networking community has to come up with new ML/DL mechanism to ensure appropriate deterrence against adversarial examples. Robustness can be introduced by incorporating approximation and fault tolerance on top of defense techniques against adversarial threats.      
    \item \textbf{Deterrence against adversarial attacks in CSON}: We have performed proactive defense against adversarial attacks by training on adversarial examples. This adversarial training procedure provides deterrence against the adversarial examples it is trained on but an unknown adversarial perturbation can evade the classifier. Table II depicts that when the classifier is trained via an adversarial training procedure, it enables the malware classifier to classify FGSM based adversarial examples correctly with $97.43\%$ accuracy after $100$ epochs but the same classifier was unable to classify BIM attacks with appropriate accuracy even after $100$ epochs of adversarial training. 
    This shows that before incorporating ML/DL techniques in support of CSON applications like routing, intrusion detection, traffic classification, malware detection, the research community needs to figure out an appropriate defense against all adversarial perturbations. The margin of error in adversarial examples classification is very narrow in networking application when compared to computer vision problems.
   
    Building deterrence against adversarial examples requires a method to improve generalization, this can be achieved via constraint objective function optimization, distributed denoising, and exploiting vicinal risk minimization instead of empirical losses. Apple Inc. \cite{moosavi2018divide} proposed a distributed denoising scheme for building deterrence against adversarial attacks for security-critical applications whereas Zhang et al. \cite{zhang2017mixup} proposed a method for improving the generalization of the ML/DL schemes which uses vicinal risk minimization rather than conventional empirical loss minimization. This procedure improves the robustness of ML/DL techniques against adversarial examples. Our experiments demonstrate that CSON are currently lacking the capability to provide appropriate defense against adversarial attacks on ML/DL techniques. 
    
    \item \textbf{Arms race between adversaries}: Our experiments also highlight that using ML/DL techniques in CSON can lead to an arms race situation between adversaries. Consequently, adversarial attacks and defense mechanisms will be in an arms race where attackers keep on dynamically changing the adversarial perturbations and defenders have to adapt accordingly. 
    
\end{itemize}
ML/DL techniques will enable future CSON but before their deployments, the research community has to figure out an effective way to deal with adversarial attacks.

\subsection{Open issues}
\begin{itemize}
    \item \textbf{Standardized datasets}: Progress in CSON largely depends upon learning from data obtained from the user, operating system, and application. Unfortunately, there does not exist a single standardized dataset for benchmarking ML/DL techniques for real-time networking applications. In order to ensure a proper utilization of ML/DL techniques with efficient deterrence against adversarial examples networking community has to come up with standardized datasets for security-critical applications. 
    \item \textbf{Learning from untapped network data}: Building deterrence in CSON against adversarial examples can be achieved by improving the generalization of ML/DL techniques. Generalization can be improved by harnessing the features from untapped networking data (network data that is recorded but not utilized in decision making) by introducing new network telemetry schemes for CSON. This can be a very promising way forward in realizing security critical CSON.  
    \item \textbf{New ML/DL mechanisms}: Conventional ML/DL techniques are very vulnerable to adversarial examples as shown in section \ref{sec4} and related defense schemes do not qualify for CSON applications. Developing new ML/DL schemes for unstructured networking data which are robust to adversarial threats is still an open avenue. Geometric and graph ML/DL techniques have the potential to solve this issue but have not yet been explored in this context. 
\end{itemize}
\section{Conclusion}
In this paper, we evaluated the feasibility of employing ML/DL techniques to realize CSON in security critical applications and their ability to defend against adversarial examples. We demonstrated that network data is highly susceptible to adversarial attacks. We also evaluated the proactive defense mechanisms to build a defense against adversarial perturbations. Our experiments demonstrate that the application of ML/DL techniques in networking can push the limits on the state-of-the-art in CSON. However, without taking into account the threat of adversarial examples, significant security risks will be a major hindrance to the deployment of these networks. 



\end{document}